\documentclass[prd,aps,twocolumn,showpacs,floats,floatfix]{revtex4}
\usepackage{graphicx}
\usepackage{amsmath}
\usepackage{physics}
\usepackage{subfigure}
\usepackage{appendix}
\usepackage[normalem]{ulem}
\usepackage[utf8]{inputenc}
\usepackage{color}

\begin{document}

\title{A morphology-independent search for gravitational wave echoes in 
data from the first and second observing runs of Advanced LIGO and Advanced Virgo}

\author{
Ka Wa Tsang$^{1,2}$,
 Archisman Ghosh$^{1}$,
 Anuradha Samajdar$^{1}$,
 Katerina Chatziioannou$^{3}$,
 Simone Mastrogiovanni$^{4}$,
 Michalis Agathos$^{5}$, 
 and Chris Van Den Broeck$^{1,2}$}

\affiliation{$^1$Nikhef -- National Institute for Subatomic Physics, 105 Science Park, 
1098 XG Amsterdam, The Netherlands \\
$^2$Van Swinderen Institute for Particle Physics and Gravity, University of Groningen, \\
Nijenborgh 4, 9747 AG Groningen, The Netherlands \\
$^3$Center for Computational Astrophysics, Flatiron Institute, 162 5th Ave, New York, NY 10010, USA \\
$^4$APC, AstroParticule et Cosmologie, Université Paris Diderot, CNRS/IN2P3, CEA/Irfu,
Observatoire de Paris, Sorbonne Paris Cité, F-75205 Paris Cedex 13, France\\
$^5$Theoretical Physics Institute, University of Jena, 07743 Jena, Germany
}

\date{\today}


\begin{abstract}

Gravitational wave echoes have been proposed as a smoking-gun signature 
of exotic compact objects with 
near-horizon structure. Recently there have been observational claims that echoes are indeed 
present in stretches of data from Advanced LIGO and Advanced Virgo 
immediately following 
gravitational wave signals from presumed binary black hole mergers, as well as a binary 
neutron star merger. In this paper we deploy a morphology-independent search algorithm 
for echoes introduced in Tsang \emph{et al.}, Phys.~Rev.~D {\bf 98}, 024023 (2018), which 
(a) is able to accurately reconstruct a possible echoes signal 
with minimal assumptions about their morphology, and
(b) computes Bayesian evidences for the hypotheses that the data contain a signal, an instrumental glitch, 
or just stationary, Gaussian noise. 
Here we apply this analysis method to all the significant events in 
the first Gravitational Wave Transient Catalog (GWTC-1), which comprises the signals 
from binary black hole and binary neutron star coalescences found during the first 
and second observing runs of Advanced LIGO and Advanced Virgo. In all cases, 
the ratios of evidences for signal versus noise and signal versus glitch do not 
rise above their respective ``background distributions" obtained from detector noise, 
the smallest $p$-value being 3\% (for event GW170823). 
Hence we find no statistically significant evidence for echoes in GWTC-1.
\end{abstract}


\maketitle


\emph{Introduction.} 
Over the past several years, the twin Advanced LIGO observatories \cite{TheLIGOScientific:2014jea} 
have been detecting gravitational wave (GW) signals from coalescences of compact binary objects on a regular 
basis \cite{Abbott:2017xlt,Abbott:2016nmj,TheLIGOScientific:2016pea,Abbott:2017vtc,Abbott:2017gyy}. 
Meanwhile Advanced Virgo \cite{TheVirgo:2014hva} has joined the global network of detectors, leading to 
further detections, including a binary neutron star inspiral \cite{TheLIGOScientific:2017qsa}. 
In the first and second observing runs 
a total of 11 detections were made, which are summarized in \cite{LIGOScientific:2018mvr}; the
latter reference will be referred to as GWTC-1 (for Gravitational Wave Transient Catalog 1).

Thanks to these observations, general relativity (GR) has been subjected to a range of tests. 
For the first time we had access to the genuinely strong-field dynamics of the theory 
\cite{Abbott:2017xlt,TheLIGOScientific:2016src,TheLIGOScientific:2016pea,Abbott:2017vtc,Abbott:2018lct}.
Possible dispersion of gravitational waves was strongly constrained, leading to stringent 
upper bounds on the mass of the graviton and on local Lorentz invariance violations  
\cite{TheLIGOScientific:2016src,Abbott:2017vtc,Abbott:2018lct}. As a next step, we want 
to probe the nature of the compact objects themselves. Based on the available data, how certain
can we be that the more massive compact objects that were observed were indeed the standard
black holes of classical, vacuum general relativity? A variety of alternative objects 
(``black hole mimickers") have been proposed; see \emph{e.g.}~\cite{Barack:2018yly} for an overview.
When such objects are part of a binary system that undergoes coalescence, anomalous effects associated with 
them
can leave an imprint upon the observed gravitational wave signal, including tidal effects 
\cite{Cardoso:2017cfl,Johnson-McDaniel:2018uvs}; dynamical friction as well as resonant 
excitations due to dark matter clouds surrounding the objects \cite{Baumann:2018vus}; 
violations of the no-hair conjecture \cite{Carullo:2018sfu,Brito:2018rfr}; and finally 
through gravitational wave ``echoes" that might follow a merger 
\cite{Cardoso:2016rao,Cardoso:2016oxy,Cardoso:2017cqb,Cardoso:2019apo}. 

In this paper we will in particular search for echoes. In the case of exotic compact objects 
that lack a horizon, ingoing gravitational waves (\emph{e.g.}~resulting from a merger) 
can reflect multiple times off effective radial potential barriers, with wave packets leaking out at set 
times and escaping to infinity. Given an exotic object 
with mass $M$ and a horizon modification with typical length scale $\ell$, the time between these echoes 
tends to be constant, and approximately equal to $\Delta t \simeq nM \log(M/\ell)$, with 
$n$ a factor that is determined by the nature of the object \cite{Cardoso:2016oxy}. 
Setting $\ell$ equal to the Planck length, for the masses involved in the binary coalescences
of GWTC-1 one can expect $\Delta t$ to range from a few to a few hundred milliseconds.

In \cite{Abedi:2016hgu,Westerweck:2017hus}, searches for echoes were presented using 
a bank of template waveforms characterized by the above $\Delta t$ as well as 
a characteristic frequency, a damping factor, and a widening factor between successive echoes.
Since then, template-based search methods were developed which explore 
the relevant parameter space in a more efficient way \cite{Lo:2018sep,Nielsen:2018lkf}, or
use more sophisticated templates \cite{Uchikata:2019frs}. A 
potential drawback here is that echo waveforms have been explicitly calculated only for 
selected exotic objects under various assumptions \cite{Cardoso:2016oxy,Cardoso:2017cqb},
and even then only in an exploratory way \cite{Mark:2017dnq}. Hence it is desirable to have
a method to search for and characterize \emph{generic} echoes, irrespective of their
detailed morphology. A model-independent search for echoes was presented in \cite{Abedi:2018npz},
but essentially assuming that individual echoes can be approximated by Dirac delta functions.
In \cite{Conklin:2017lwb} a search was performed by looking for coherent excess power in 
a succession of windows in time or frequency.   
In this paper we instead employ the framework developed in \cite{Tsang:2018uie} based on the
\texttt{BayesWave} algorithm \cite{Cornish:2014kda,Littenberg:2014oda} which can be used to 
not only detect but also reconstruct and characterize echo signals of an 
\emph{a priori} unknown form.\\ 

\emph{Method.}
The method we use was extensively described in \cite{Tsang:2018uie}; here we only give an overview and
then describe how it was applied to data from GWTC-1. 
We model the detector data ${\bf s}$ as 
\begin{equation}
{\bf s} = {\bf R} \ast {\bf h} + {\bf g} + {\bf n_g}, 
\label{models}
\end{equation}
where ${\bf R}$ is the response 
of the network to gravitational waves, ${\bf h}$ is the potential signal that is coherent across the detectors, 
${\bf g}$ denotes possible instrumental transients or glitches, and ${\bf n_g}$ is a contribution from stationary, Gaussian noise.
The signal ${\bf h}$ and the glitches ${\bf g}$ can both be decomposed in terms of
a set of basis functions, and Bayesian evidences can be obtained for the hypotheses that either are 
present in the data. 
An important difference between signals and glitches is that the former will be present in 
the data of all the detectors in a coherent way (taking into account the different antenna responses), 
whereas the latter will manifest themselves incoherently. If the data contain a coherent signal, then typically
a smaller number of basis functions will be needed to reconstruct it than to reconstruct 
incoherent glitches, so that the glitch model incurs an Occam penalty. At the same time, 
a reconstruction of the shape of the signal is obtained from the corresponding superposition
of basis functions. For our purposes a natural choice for the basis functions is a 
``train" of sine-Gaussians. Individual sine-Gaussians are characterized by an amplitude $A$, 
a central frequency 
$f_0$, a damping time $\tau$, and a reference phase $\phi_0$; the train of sine-Gaussians
as a whole also involves a central time $t_0$ of the first sine-Gaussian, 
a time $\Delta t$ between successive
sine-Gaussians, and in going from one sine-Gaussian to the next also a relative phase shift
$\Delta\phi$, an amplitude damping factor $\gamma$, and a widening factor $w$. 
Although there is no expectation that real echoes signals would resemble
any one of these ``generalized wavelets", it is reasonable to assume that \emph{superpositions} of 
them will be able to catch a wide variety of physical echoes waveforms and, if the first few
echoes are sufficiently loud, provide an adequate reconstruction. 
Finally, the noise model ${\bf n_g}$
consists of colored Gaussian noise, the power spectral density of which is computed using 
a combination of smooth spline curves and a collection of Lorentzians to 
fit sharp spectral lines \cite{Littenberg:2014oda}. For each of the three models, the relevant 
parameter space is sampled over using a reversible jump Markov chain Monte Carlo
algorithm, in which the number of generalized wavelets is also allowed to 
vary freely. Evidences for the three hypotheses 
are estimated through thermodynamic integration, yielding Bayes factors 
$B_{S/N}$ and $B_{S/G}$ for, respectively, the signal versus noise and 
signal versus glitch models \cite{Cornish:2014kda}. This allows us to not only perform model 
selection,
but also to reconstruct and characterize the signal. The algorithm is
also applied to many stretches of detector noise, leading to a 
background distribution for $B_{S/N}$ and $B_{S/G}$ which can then be
used to assess the statistical significance of potential echoes signals in 
the usual way. For more details we refer to \cite{Tsang:2018uie}. 

In analyzing the stretches of data immediately preceding the events in GWTC-1 (for
background calculation) or immediately after them (to search for echoes), we need to 
choose priors. We take $f_0$ to be uniform in the interval $[30, 1024]$ Hz 
(respectively the lower cut-off frequency and half the sampling rate of the analysis), and the quality 
factor $Q = 2\pi f_0 \tau \in [2,40]$ uniformly, so that $\tau$ takes values roughly between
$3 \times 10^{-4}$ s and $0.2$ s, consistent with time scales set by the masses involved
in the events. We let $\phi_0$ be uniform in $[0,2\pi]$. 
The prior on $A$ is based on signal-to-noise ratio as in \cite{Cornish:2014kda}. 
We take uniform  priors $\Delta t \in [0,0.7]$ s, $\Delta\phi \in [0,2\pi]$,
$\gamma \in [0,1]$, and $w \in [1,2]$. For definiteness, each generalized wavelet  
contains five sine-Gaussians. 
We also need to specify a prior for the central 
time of the first sine-Gaussian in a generalized wavelet. Here we want to 
start analyzing at a time that is safely beyond the plausible duration of the 
ringdown of the remnant object. Let $t_{\rm event}$ be the
arrival time for a given binary coalescence event as given in \cite{LIGOScientific:2018mvr}; then
we take $t_0$ to be uniform in 
$[t_{\rm event} + 4\tau_{220}, t_{\rm event} + 4\tau_{220} + 0.5\,\mbox{s}]$. The value for 
$\tau_{220}$ is a conservatively long estimate for the decay time of the $220$ mode in the
ringdown, using the fitting formula $\tau_{220}(M_f,a_f,z)$ of \cite{Berti:2005ys}, where 
for the final mass $M_f$, the final spin $a_f$, and the redshift $z$ we take values at the
upper bounds of the 90\% confidence intervals listed in \cite{LIGOScientific:2018mvr}; typically
this comes to a few milliseconds. We note that our choices for parameter prior ranges, 
though pertaining to generalized wavelet decompositions rather than waveform templates, 
include the corresponding values for 
$t_0$, $\gamma$, and $A$ at which the template-based analysis of \cite{Abedi:2016hgu} claimed 
tentative evidence for echoes related to GW150914, GW151012, and GW151226.  

To construct background distributions for the log Bayes factors 
$\log B_{S/N}$ and $\log B_{S/G}$, we use stretches of 
data \emph{preceding}
each coalescence event in GWTC-1 in the following way. In the interval between 1050 s and 250 s before 
the GPS time of 
a binary coalescence trigger as given in \cite{LIGOScientific:2018mvr}, we define 100 sub-intervals
of 8 s each. (No signal 
in GWTC-1 will have been in the detectors' sensitive frequency band for more than 250 s, 
hence these intervals should effectively contain noise only.) For each of these intervals
we compute $\log B_{S/N}$ and $\log B_{S/G}$, where the priors for the parameters of the
generalized wavelets are as explained above; the values for $t_{\rm event}$ are chosen 
to be at the start of each interval. 
The log Bayes factors from times preceding all the events that were seen in two detectors obtained in 
this way 
are put into histograms, and the same is done separately for log Bayes factors from times 
preceding all the 3-detector events. These histograms are normalized, and 
smoothened using Gaussian kernel 
density estimates to obtain approximate probability distributions for 
$\log B_{S/N}$ and $\log B_{S/G}$ in the absence of echoes signals.

\begin{figure}[h!]
 \includegraphics[width=0.45\textwidth]{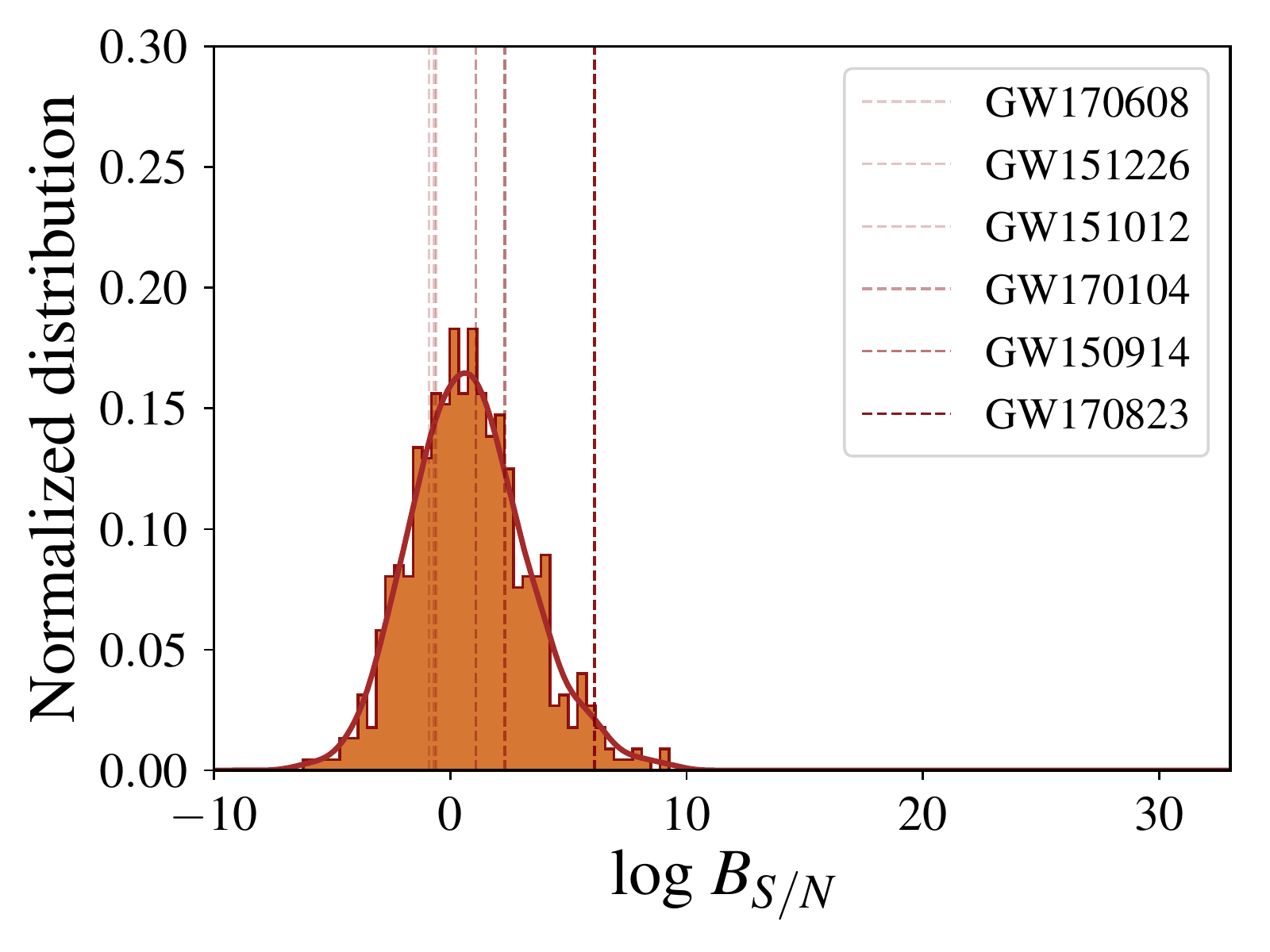}
 \includegraphics[width=0.45\textwidth]{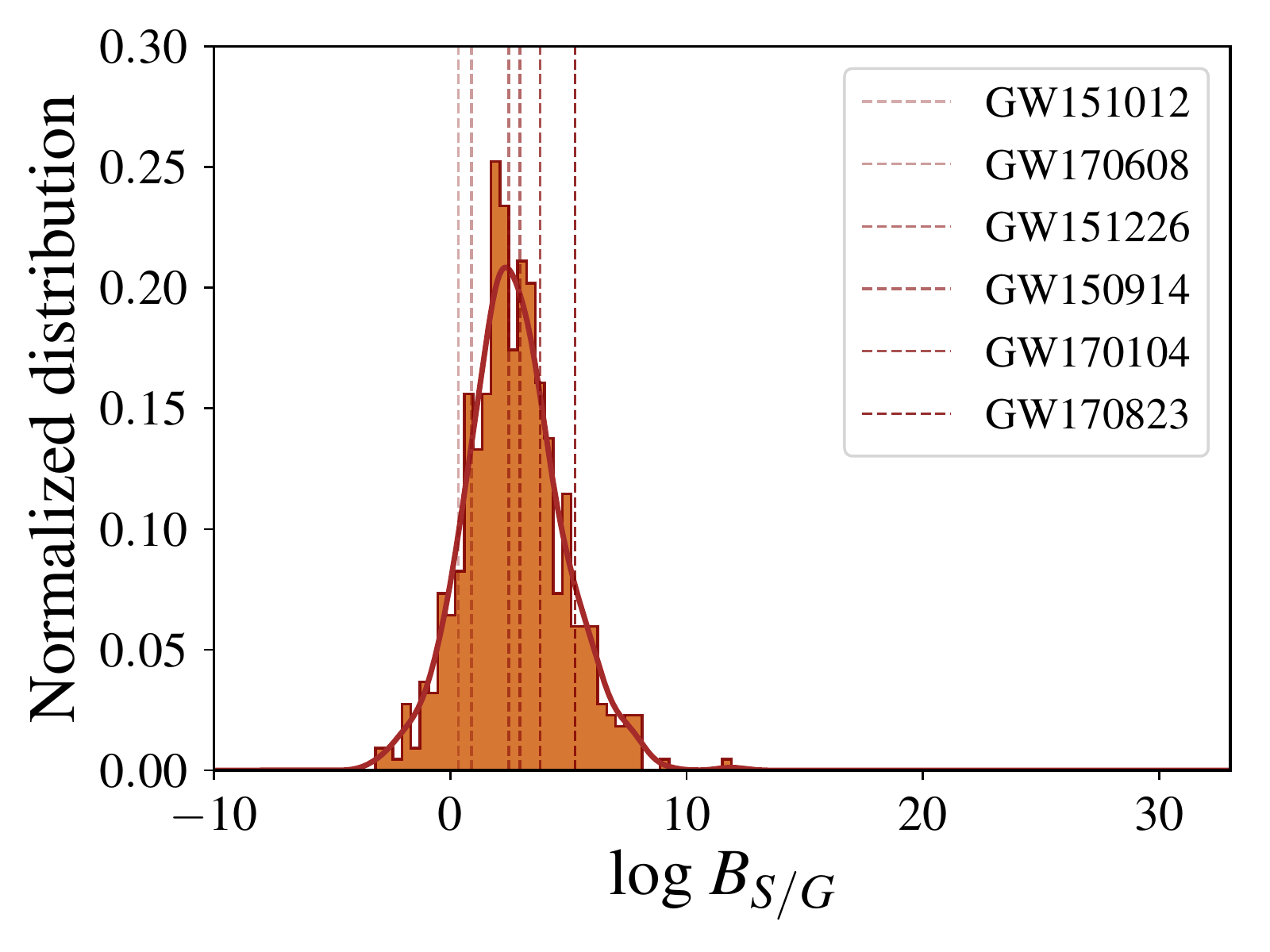}
 \caption{Background distributions (orange histograms, smoothened in brown) and foreground (vertical dashed lines, shaded by their values and labelled left to right) for the 
 log Bayes factors for
 signal versus noise $\log B_{S/N}$ (top) and signal versus glitch $\log B_{S/G}$ (bottom),
 for the 2-detector events of GWTC-1. The 
 associated $p$-values can be found in Table~\ref{tab:foreground_2ifo}.}
 \label{fig:bkgdistr_2ifo}
\end{figure}

\begin{figure}[h!]
 \includegraphics[width=0.45\textwidth]{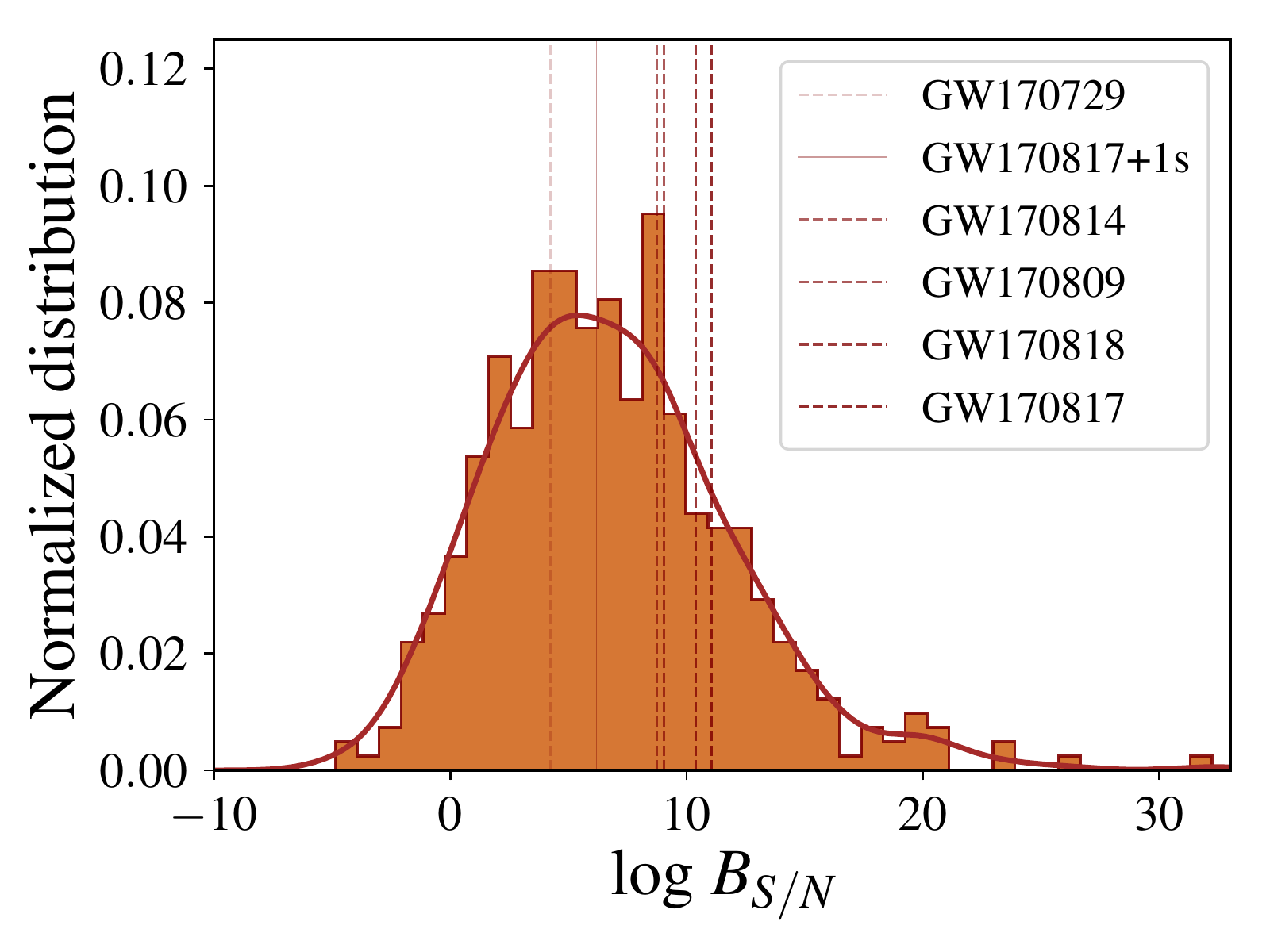}
 \includegraphics[width=0.45\textwidth]{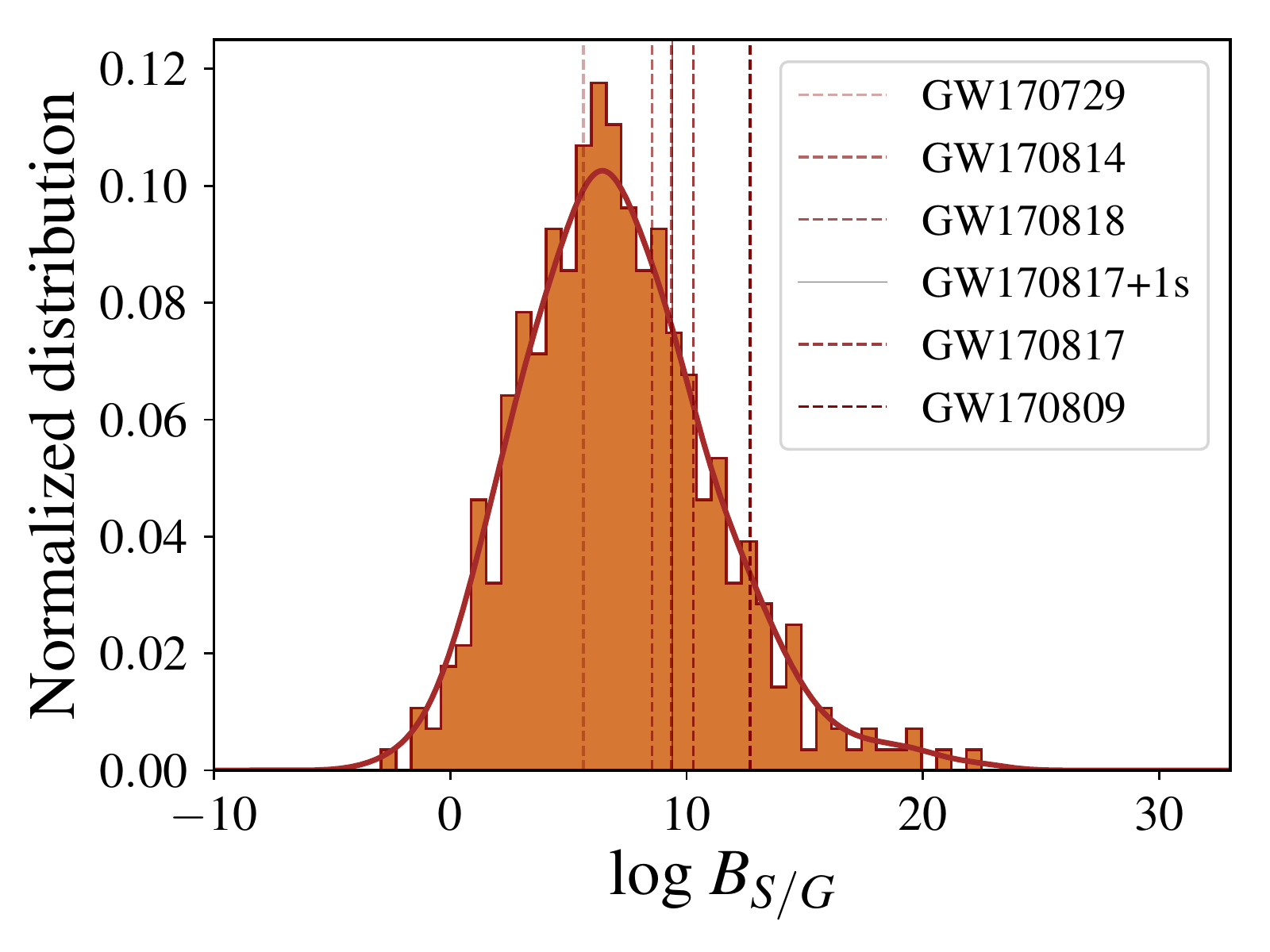}
 \caption{The same as in Fig.~\ref{fig:bkgdistr_2ifo}, but now for the 3-detector events
 of GWTC-1. The thin solid lines in each of the two panels are for an analysis 
of GW170817 in which the prior range for the time of the first echo was centered
at 1.0 s after the event time, where \cite{Abedi:2018npz} claimed tentative evidence for an echo. 
  For the 
  associated $p$-values, 
 see Table~\ref{tab:foreground_3ifo}.}
 \label{fig:bkgdistr_3ifo}
\end{figure}

\begin{table}[]
\begin{tabular}{lllllllll}
Event & \vline & $\log B_{S/N}$ & \vline & $p_{S/N}$ & \vline & $\log B_{S/G}$ & \vline & $p_{S/G}$  \\
\hline
\hline
GW150914 & \vline & 2.32  & \vline & 0.26 & \vline & 2.95 & \vline & 0.43 \\
GW151012 & \vline & -0.59 & \vline & 0.70 & \vline & 0.35 & \vline & 0.88 \\
GW151226 & \vline & -0.67 & \vline & 0.72 & \vline & 2.48 & \vline & 0.53 \\
GW170104 & \vline & 1.09  & \vline & 0.44 & \vline & 3.80 & \vline & 0.28 \\
GW170608 & \vline & -0.90 & \vline & 0.75 & \vline & 0.90 & \vline & 0.82 \\
GW170823 & \vline & 6.11  & \vline & 0.03 & \vline & 5.29 & \vline & 0.11  \\
Combined & \vline &       & \vline & 0.34 & \vline &      & \vline & 0.57           
\end{tabular}
\caption{Log Bayes factors for signal versus noise and signal versus glitch, 
and the corresponding $p$-values,  
for events seen in two detectors. The bottom row contains the 
combined $p$-values for all these events together.}
\label{tab:foreground_2ifo}
\end{table}

\begin{table}[]
\begin{tabular}{lllllllll}
Event & \vline & $\log B_{S/N}$ & \vline & $p_{S/N}$ & \vline & $\log B_{S/G}$ & \vline & $p_{S/G}$  \\
\hline
\hline
GW170729 & \vline & 4.24  & \vline & 0.67 & \vline & 5.64  & \vline & 0.62 \\ 
GW170809 & \vline & 9.05  & \vline & 0.31 & \vline & 12.69 & \vline & 0.09 \\             
GW170814 & \vline & 8.75  & \vline & 0.33 & \vline & 8.54  & \vline & 0.34 \\ 
GW170817 & \vline & 11.05 & \vline & 0.19 & \vline & 10.30 & \vline & 0.20  \\ 
GW170817+1s & \vline & 6.19  & \vline & 0.52 & \vline & 9.39 & \vline & 0.27 \\ 
GW170818 & \vline & 10.39 & \vline & 0.23 & \vline & 9.36 & \vline & 0.27 \\ 
Combined & \vline &       & \vline & 0.47 & \vline &      & \vline & 0.22                 
\end{tabular}
\caption{The same as in Table \ref{tab:foreground_3ifo}, but now for the events
that were seen in three detectors. In the case of GW170817 we also include results 
for which the prior range for the time of the first echo was centered
at 1.0 s after the event time, where \cite{Abedi:2018npz} claimed tentative evidence for an echo.
The combined $p$-values take the latter prior choice for this particular event.}
\label{tab:foreground_3ifo}
\end{table}

Finally,     
we calculate log Bayes factors for times \emph{following} the coalescence events
of GWTC-1, using the same priors, but now setting $t_{\rm event}$ to the arrival times 
for the events given in \cite{LIGOScientific:2018mvr}. Considering $\log B_{S/N}$ or $\log B_{S/G}$
and the relevant number of detectors, the normalized, smoothened background distributions 
$\mathcal{P}(\log B)$ are used to compute $p$-values:
\begin{equation}
p = 1 - \int_{-\infty}^{\log B} \mathcal{P}(x)\,dx.
\end{equation}
Combined $p$-values from all the events are obtained using Fisher's prescription
\cite{fisher_method}. Given individual 
$p$-values $p_i$, $i = 1, \ldots, N$, one defines
\begin{equation}
S = -2 \sum_{i=1}^N \log(p_i), 
\end{equation}
and the combined $p$-value is calculated as
\begin{equation}
p_{\rm comb} = 1 - \int_0^S \chi^2_{2N}(x)\,dx,
\end{equation}
where $\chi^2_{2N}$ is the chi-squared distribution with $2N$ degrees of freedom.\\

\newpage
\begin{figure*}[h!]
  \includegraphics[width=17cm,trim={0 8cm 0 0},clip]{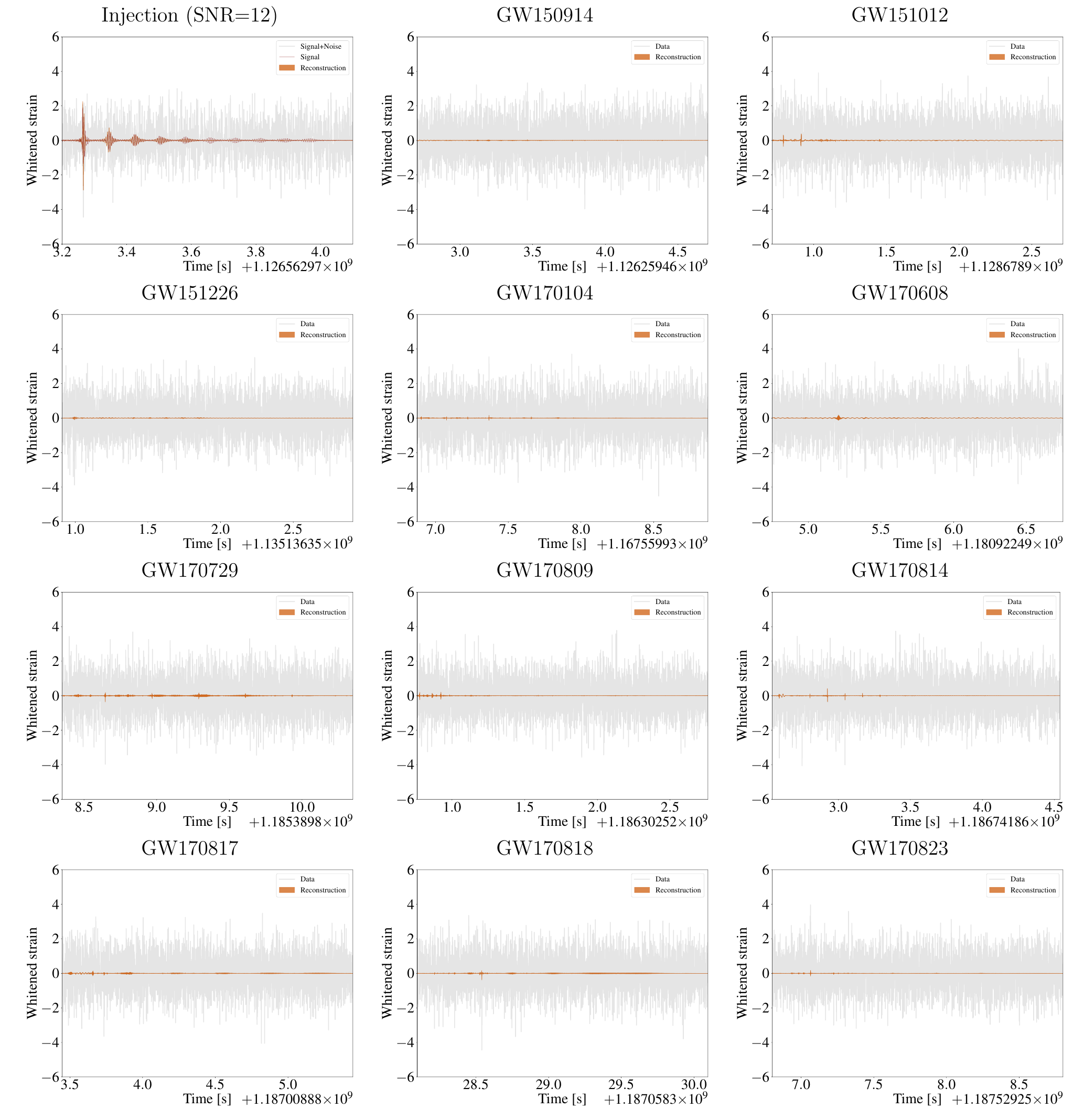}
 \caption{Stretches of whitened data (gray) and signal reconstructions (red) 
 for a simulated echoes signal as described in the main text (top left panel) and for 
 data immediately after the events in GWTC-1. In the case of GW170817, 
 the first echo is searched for in an interval centered at 1.0 s after the 
 event time. In all cases the event GPS time corresponds to the left bound of the panel.}
 \label{fig:reconstructions}
\end{figure*}

\emph{Results and discussion.}
Figures~\ref{fig:bkgdistr_2ifo} and \ref{fig:bkgdistr_3ifo}
show background 
and foreground for $\log B_{S/N}$ and $\log B_{S/G}$ in the case of, respectively, 
2-detector and 3-detector signals, and in Tables \ref{tab:foreground_2ifo} and \ref{tab:foreground_3ifo}
we list the specific log Bayes factors for these cases, as well as associated $p$-values. 
All foreground results 
are in the support of the relevant background distributions. For signal versus noise, 
the smallest $p$-value is 3\% (the case of GW170823), whereas for signal versus glitch the $p$-values do not
go below 9\% (see GW170809). In summary, we find 
no statistically significant evidence for echoes in GWTC-1. For 
the binary black hole observations in particular, this statement is in agreement with the template-based
searches in \cite{Westerweck:2017hus,Nielsen:2018lkf,Uchikata:2019frs}. (Note that 
a \emph{quantitative} comparison of $p$-values is hard to make, because of the very specific signal shapes 
that are assumed in the latter analyses.)
  
Our results in Fig.~\ref{fig:bkgdistr_3ifo} and Table \ref{tab:foreground_3ifo} include 
the binary neutron star inspiral GW170817, analyzed in the same manner
as the binary black hole merger signals. 
In \cite{Abbott:2018wiz}, an analysis using the original
\texttt{BayesWave} algorithm of \cite{Cornish:2014kda,Littenberg:2014oda} (\emph{i.e.}~employing
wavelets that are simple sine-Gaussians) yielded no evidence for a post-merger signal.
Using our generalized wavelets, we obtain $\log B_{S/N} = 11.05$ and 
$\log B_{S/G} = 10.30$, both consistent with background. Hence in particular we do not find 
evidence for an echoes-like post-merger signal either, at least not up to $\lesssim 0.5$ s
after the event's GPS time. 
In \cite{Abedi:2018npz}, tentative evidence was claimed for echoes starting at 
$t_0 = t_{\rm event} + 1.0$ s. Re-analyzing with the same priors as above but this time 
$t_0 \in [t_{\rm event} + 0.75\,\mbox{s}, t_{\rm event} + 1.25\,\mbox{s}]$, we find
$\log B_{S/N} = 6.19$ and $\log B_{S/G} = 9.39$, both of which are consistent with their respective
background distributions.  
 Hence also when the time of the first echo is in this time interval we find 
no significant evidence for echoes. 
That said, we explicitly note that in the case of a black hole
resulting from a binary neutron star merger of total mass 
$\sim 2.7\,M_\odot$ \cite{Abbott:2018wiz}, we expect the dominant ringdown frequency and hence the
central frequency $f_0$ of the echoes to be above 6000 Hz, 
\emph{i.e.}~above our prior upper bound, but also much beyond  
the detectors' frequency reach for plausible energies emitted \cite{Abbott:2017dke}. 
Foreground and background analyses with a 
correspondingly high frequency range are left for future work.
 
Finally, in Fig.~\ref{fig:reconstructions} we show signal reconstructions
(medians and 90\% credible intervals) in terms
of generalized wavelets for all the GWTC-1 events. For illustration purposes we also include 
the reconstruction of a simulated echoes waveform following the inspiral of a particle in a 
Schwarzschild
spacetime with Neumann reflective boundary conditions just outside of where the 
horizon would have been, at mass ratio $q = 100$ \cite{Khanna:2016yow,Price:2017cjr}. 
The simulated signal was embedded
into detector noise at a signal-to-noise ratio (SNR) of 12, roughly
corresponding to the SNR in the ringdown part of GW150914, had it been observed with 
Advanced LIGO sensitivity of the second observing run. In all cases the whitened 
raw data is shown, along with the whitened signal reconstruction. We include these 
reconstructions for completeness; our main results are the ones in Figs.~\ref{fig:bkgdistr_2ifo}, 
\ref{fig:bkgdistr_3ifo} and Tables \ref{tab:foreground_2ifo}, \ref{tab:foreground_3ifo}. Nevertheless, 
the reconstruction plots are visually consistent with our core results.





\section*{Acknowledgments}
The authors thank Vitor Cardoso, Gaurav Khanna, Alex Nielsen, and Paolo Pani for helpful discussions. 
K.W.T., A.G., A.S., and C.V.D.B.~are supported by the 
research programme of the Netherlands Organisation for Scientific Research (NWO). 
S.M.~was supported by a STSM grant from 
COST Action CA16104. This research has made use of data obtained from the Gravitational Wave Open 
Science Center (https://www.gw-openscience.org), a service of LIGO Laboratory, the LIGO Scientic 
Collaboration, and the Virgo Collaboration.


\bibliographystyle{apsrev}
\bibliography{paper}

\begin{thebibliography}{38}
\expandafter\ifx\csname natexlab\endcsname\relax\def\natexlab#1{#1}\fi
\expandafter\ifx\csname bibnamefont\endcsname\relax
  \def\bibnamefont#1{#1}\fi
\expandafter\ifx\csname bibfnamefont\endcsname\relax
  \def\bibfnamefont#1{#1}\fi
\expandafter\ifx\csname citenamefont\endcsname\relax
  \def\citenamefont#1{#1}\fi
\expandafter\ifx\csname url\endcsname\relax
  \def\url#1{\texttt{#1}}\fi
\expandafter\ifx\csname urlprefix\endcsname\relax\def\urlprefix{URL }\fi
\providecommand{\bibinfo}[2]{#2}
\providecommand{\eprint}[2][]{\url{#2}}

\bibitem[{\citenamefont{Aasi et~al.}(2015)}]{TheLIGOScientific:2014jea}
\bibinfo{author}{\bibfnamefont{J.}~\bibnamefont{Aasi}} \bibnamefont{et~al.}
  (\bibinfo{collaboration}{LIGO Scientific}),
  \bibinfo{journal}{Class.Quant.Grav.} \textbf{\bibinfo{volume}{32}},
  \bibinfo{pages}{074001} (\bibinfo{year}{2015}), \eprint{1411.4547}.

\bibitem[{\citenamefont{Abbott}(2017)}]{Abbott:2017xlt}
\bibinfo{author}{\bibfnamefont{B.~P.} \bibnamefont{Abbott}}
  (\bibinfo{year}{2017}), pp. \bibinfo{pages}{291--311}.

\bibitem[{\citenamefont{Abbott et~al.}(2016{\natexlab{a}})}]{Abbott:2016nmj}
\bibinfo{author}{\bibfnamefont{B.~P.} \bibnamefont{Abbott}}
  \bibnamefont{et~al.} (\bibinfo{collaboration}{Virgo, LIGO Scientific}),
  \bibinfo{journal}{Phys. Rev. Lett.} \textbf{\bibinfo{volume}{116}},
  \bibinfo{pages}{241103} (\bibinfo{year}{2016}{\natexlab{a}}),
  \eprint{1606.04855}.

\bibitem[{\citenamefont{Abbott
  et~al.}(2016{\natexlab{b}})}]{TheLIGOScientific:2016pea}
\bibinfo{author}{\bibfnamefont{B.~P.} \bibnamefont{Abbott}}
  \bibnamefont{et~al.} (\bibinfo{collaboration}{Virgo, LIGO Scientific}),
  \bibinfo{journal}{Phys. Rev.} \textbf{\bibinfo{volume}{X6}},
  \bibinfo{pages}{041015} (\bibinfo{year}{2016}{\natexlab{b}}),
  \eprint{1606.04856}.

\bibitem[{\citenamefont{Abbott et~al.}(2017{\natexlab{a}})}]{Abbott:2017vtc}
\bibinfo{author}{\bibfnamefont{B.~P.} \bibnamefont{Abbott}}
  \bibnamefont{et~al.} (\bibinfo{collaboration}{VIRGO, LIGO Scientific}),
  \bibinfo{journal}{Phys. Rev. Lett.} \textbf{\bibinfo{volume}{118}},
  \bibinfo{pages}{221101} (\bibinfo{year}{2017}{\natexlab{a}}),
  \eprint{1706.01812}.

\bibitem[{\citenamefont{Abbott et~al.}(2017{\natexlab{b}})}]{Abbott:2017gyy}
\bibinfo{author}{\bibfnamefont{B.~P.} \bibnamefont{Abbott}}
  \bibnamefont{et~al.} (\bibinfo{collaboration}{Virgo, LIGO Scientific}),
  \bibinfo{journal}{Astrophys. J.} \textbf{\bibinfo{volume}{851}},
  \bibinfo{pages}{L35} (\bibinfo{year}{2017}{\natexlab{b}}),
  \eprint{1711.05578}.

\bibitem[{\citenamefont{Acernese et~al.}(2015)}]{TheVirgo:2014hva}
\bibinfo{author}{\bibfnamefont{F.}~\bibnamefont{Acernese}} \bibnamefont{et~al.}
  (\bibinfo{collaboration}{Virgo}), \bibinfo{journal}{Class. Quant. Grav.}
  \textbf{\bibinfo{volume}{32}}, \bibinfo{pages}{024001}
  (\bibinfo{year}{2015}), \eprint{1408.3978}.

\bibitem[{\citenamefont{Abbott
  et~al.}(2017{\natexlab{c}})}]{TheLIGOScientific:2017qsa}
\bibinfo{author}{\bibfnamefont{B.}~\bibnamefont{Abbott}} \bibnamefont{et~al.}
  (\bibinfo{collaboration}{Virgo, LIGO Scientific}), \bibinfo{journal}{Phys.
  Rev. Lett.} \textbf{\bibinfo{volume}{119}}, \bibinfo{pages}{161101}
  (\bibinfo{year}{2017}{\natexlab{c}}), \eprint{1710.05832}.

\bibitem[{\citenamefont{Abbott
  et~al.}(2018{\natexlab{a}})}]{LIGOScientific:2018mvr}
\bibinfo{author}{\bibfnamefont{B.~P.} \bibnamefont{Abbott}}
  \bibnamefont{et~al.} (\bibinfo{collaboration}{LIGO Scientific, Virgo})
  (\bibinfo{year}{2018}{\natexlab{a}}), \eprint{1811.12907}.

\bibitem[{\citenamefont{Abbott
  et~al.}(2016{\natexlab{c}})}]{TheLIGOScientific:2016src}
\bibinfo{author}{\bibfnamefont{B.~P.} \bibnamefont{Abbott}}
  \bibnamefont{et~al.} (\bibinfo{collaboration}{Virgo, LIGO Scientific}),
  \bibinfo{journal}{Phys. Rev. Lett.} \textbf{\bibinfo{volume}{116}},
  \bibinfo{pages}{221101} (\bibinfo{year}{2016}{\natexlab{c}}),
  \eprint{1602.03841}.

\bibitem[{\citenamefont{Abbott et~al.}(2018{\natexlab{b}})}]{Abbott:2018lct}
\bibinfo{author}{\bibfnamefont{B.~P.} \bibnamefont{Abbott}}
  \bibnamefont{et~al.} (\bibinfo{collaboration}{LIGO Scientific, Virgo})
  (\bibinfo{year}{2018}{\natexlab{b}}), \eprint{1811.00364}.

\bibitem[{\citenamefont{Barack et~al.}(2018)}]{Barack:2018yly}
\bibinfo{author}{\bibfnamefont{L.}~\bibnamefont{Barack}} \bibnamefont{et~al.}
  (\bibinfo{year}{2018}), \eprint{1806.05195}.

\bibitem[{\citenamefont{Cardoso et~al.}(2017)\citenamefont{Cardoso, Franzin,
  Maselli, Pani, and Raposo}}]{Cardoso:2017cfl}
\bibinfo{author}{\bibfnamefont{V.}~\bibnamefont{Cardoso}},
  \bibinfo{author}{\bibfnamefont{E.}~\bibnamefont{Franzin}},
  \bibinfo{author}{\bibfnamefont{A.}~\bibnamefont{Maselli}},
  \bibinfo{author}{\bibfnamefont{P.}~\bibnamefont{Pani}}, \bibnamefont{and}
  \bibinfo{author}{\bibfnamefont{G.}~\bibnamefont{Raposo}},
  \bibinfo{journal}{Phys. Rev.} \textbf{\bibinfo{volume}{D95}},
  \bibinfo{pages}{084014} (\bibinfo{year}{2017}), \bibinfo{note}{[Addendum:
  Phys. Rev.D95,no.8,089901(2017)]}, \eprint{1701.01116}.

\bibitem[{\citenamefont{Johnson-Mcdaniel
  et~al.}(2018)\citenamefont{Johnson-Mcdaniel, Mukherjee, Kashyap, Ajith,
  Del~Pozzo, and Vitale}}]{Johnson-McDaniel:2018uvs}
\bibinfo{author}{\bibfnamefont{N.~K.} \bibnamefont{Johnson-Mcdaniel}},
  \bibinfo{author}{\bibfnamefont{A.}~\bibnamefont{Mukherjee}},
  \bibinfo{author}{\bibfnamefont{R.}~\bibnamefont{Kashyap}},
  \bibinfo{author}{\bibfnamefont{P.}~\bibnamefont{Ajith}},
  \bibinfo{author}{\bibfnamefont{W.}~\bibnamefont{Del~Pozzo}},
  \bibnamefont{and} \bibinfo{author}{\bibfnamefont{S.}~\bibnamefont{Vitale}}
  (\bibinfo{year}{2018}), \eprint{1804.08026}.

\bibitem[{\citenamefont{Baumann et~al.}(2019)\citenamefont{Baumann, Chia, and
  Porto}}]{Baumann:2018vus}
\bibinfo{author}{\bibfnamefont{D.}~\bibnamefont{Baumann}},
  \bibinfo{author}{\bibfnamefont{H.~S.} \bibnamefont{Chia}}, \bibnamefont{and}
  \bibinfo{author}{\bibfnamefont{R.~A.} \bibnamefont{Porto}},
  \bibinfo{journal}{Phys. Rev.} \textbf{\bibinfo{volume}{D99}},
  \bibinfo{pages}{044001} (\bibinfo{year}{2019}), \eprint{1804.03208}.

\bibitem[{\citenamefont{Carullo et~al.}(2018)}]{Carullo:2018sfu}
\bibinfo{author}{\bibfnamefont{G.}~\bibnamefont{Carullo}} \bibnamefont{et~al.},
  \bibinfo{journal}{Phys. Rev.} \textbf{\bibinfo{volume}{D98}},
  \bibinfo{pages}{104020} (\bibinfo{year}{2018}), \eprint{1805.04760}.

\bibitem[{\citenamefont{Brito et~al.}(2018)\citenamefont{Brito, Buonanno, and
  Raymond}}]{Brito:2018rfr}
\bibinfo{author}{\bibfnamefont{R.}~\bibnamefont{Brito}},
  \bibinfo{author}{\bibfnamefont{A.}~\bibnamefont{Buonanno}}, \bibnamefont{and}
  \bibinfo{author}{\bibfnamefont{V.}~\bibnamefont{Raymond}},
  \bibinfo{journal}{Phys. Rev.} \textbf{\bibinfo{volume}{D98}},
  \bibinfo{pages}{084038} (\bibinfo{year}{2018}), \eprint{1805.00293}.

\bibitem[{\citenamefont{Cardoso
  et~al.}(2016{\natexlab{a}})\citenamefont{Cardoso, Franzin, and
  Pani}}]{Cardoso:2016rao}
\bibinfo{author}{\bibfnamefont{V.}~\bibnamefont{Cardoso}},
  \bibinfo{author}{\bibfnamefont{E.}~\bibnamefont{Franzin}}, \bibnamefont{and}
  \bibinfo{author}{\bibfnamefont{P.}~\bibnamefont{Pani}},
  \bibinfo{journal}{Phys. Rev. Lett.} \textbf{\bibinfo{volume}{116}},
  \bibinfo{pages}{171101} (\bibinfo{year}{2016}{\natexlab{a}}),
  \bibinfo{note}{[Erratum: Phys. Rev. Lett.117,no.8,089902(2016)]},
  \eprint{1602.07309}.

\bibitem[{\citenamefont{Cardoso
  et~al.}(2016{\natexlab{b}})\citenamefont{Cardoso, Hopper, Macedo, Palenzuela,
  and Pani}}]{Cardoso:2016oxy}
\bibinfo{author}{\bibfnamefont{V.}~\bibnamefont{Cardoso}},
  \bibinfo{author}{\bibfnamefont{S.}~\bibnamefont{Hopper}},
  \bibinfo{author}{\bibfnamefont{C.~F.~B.} \bibnamefont{Macedo}},
  \bibinfo{author}{\bibfnamefont{C.}~\bibnamefont{Palenzuela}},
  \bibnamefont{and} \bibinfo{author}{\bibfnamefont{P.}~\bibnamefont{Pani}},
  \bibinfo{journal}{Phys. Rev.} \textbf{\bibinfo{volume}{D94}},
  \bibinfo{pages}{084031} (\bibinfo{year}{2016}{\natexlab{b}}),
  \eprint{1608.08637}.

\bibitem[{\citenamefont{Cardoso and Pani}(2017)}]{Cardoso:2017cqb}
\bibinfo{author}{\bibfnamefont{V.}~\bibnamefont{Cardoso}} \bibnamefont{and}
  \bibinfo{author}{\bibfnamefont{P.}~\bibnamefont{Pani}},
  \bibinfo{journal}{Nat. Astron.} \textbf{\bibinfo{volume}{1}},
  \bibinfo{pages}{586} (\bibinfo{year}{2017}), \eprint{1709.01525}.

\bibitem[{\citenamefont{Cardoso et~al.}(2019)\citenamefont{Cardoso, Foit, and
  Kleban}}]{Cardoso:2019apo}
\bibinfo{author}{\bibfnamefont{V.}~\bibnamefont{Cardoso}},
  \bibinfo{author}{\bibfnamefont{V.~F.} \bibnamefont{Foit}}, \bibnamefont{and}
  \bibinfo{author}{\bibfnamefont{M.}~\bibnamefont{Kleban}}
  (\bibinfo{year}{2019}), \eprint{1902.10164}.

\bibitem[{\citenamefont{Abedi et~al.}(2017)\citenamefont{Abedi, Dykaar, and
  Afshordi}}]{Abedi:2016hgu}
\bibinfo{author}{\bibfnamefont{J.}~\bibnamefont{Abedi}},
  \bibinfo{author}{\bibfnamefont{H.}~\bibnamefont{Dykaar}}, \bibnamefont{and}
  \bibinfo{author}{\bibfnamefont{N.}~\bibnamefont{Afshordi}},
  \bibinfo{journal}{Phys. Rev.} \textbf{\bibinfo{volume}{D96}},
  \bibinfo{pages}{082004} (\bibinfo{year}{2017}), \eprint{1612.00266}.

\bibitem[{\citenamefont{Westerweck et~al.}(2017)\citenamefont{Westerweck,
  Nielsen, Fischer-Birnholtz, Cabero, Capano, Dent, Krishnan, Meadors, and
  Nitz}}]{Westerweck:2017hus}
\bibinfo{author}{\bibfnamefont{J.}~\bibnamefont{Westerweck}},
  \bibinfo{author}{\bibfnamefont{A.}~\bibnamefont{Nielsen}},
  \bibinfo{author}{\bibfnamefont{O.}~\bibnamefont{Fischer-Birnholtz}},
  \bibinfo{author}{\bibfnamefont{M.}~\bibnamefont{Cabero}},
  \bibinfo{author}{\bibfnamefont{C.}~\bibnamefont{Capano}},
  \bibinfo{author}{\bibfnamefont{T.}~\bibnamefont{Dent}},
  \bibinfo{author}{\bibfnamefont{B.}~\bibnamefont{Krishnan}},
  \bibinfo{author}{\bibfnamefont{G.}~\bibnamefont{Meadors}}, \bibnamefont{and}
  \bibinfo{author}{\bibfnamefont{A.~H.} \bibnamefont{Nitz}}
  (\bibinfo{year}{2017}), \eprint{1712.09966}.

\bibitem[{\citenamefont{Lo et~al.}(2018)\citenamefont{Lo, Li, and
  Weinstein}}]{Lo:2018sep}
\bibinfo{author}{\bibfnamefont{R.~K.~L.} \bibnamefont{Lo}},
  \bibinfo{author}{\bibfnamefont{T.~G.~F.} \bibnamefont{Li}}, \bibnamefont{and}
  \bibinfo{author}{\bibfnamefont{A.~J.} \bibnamefont{Weinstein}}
  (\bibinfo{year}{2018}), \eprint{1811.07431}.

\bibitem[{\citenamefont{Nielsen et~al.}(2018)\citenamefont{Nielsen, Capano,
  Birnholtz, and Westerweck}}]{Nielsen:2018lkf}
\bibinfo{author}{\bibfnamefont{A.~B.} \bibnamefont{Nielsen}},
  \bibinfo{author}{\bibfnamefont{C.~D.} \bibnamefont{Capano}},
  \bibinfo{author}{\bibfnamefont{O.}~\bibnamefont{Birnholtz}},
  \bibnamefont{and}
  \bibinfo{author}{\bibfnamefont{J.}~\bibnamefont{Westerweck}}
  (\bibinfo{year}{2018}), \eprint{1811.04904}.

\bibitem[{\citenamefont{Uchikata et~al.}(2019)\citenamefont{Uchikata, Nakano,
  Narikawa, Sago, Tagoshi, and Tanaka}}]{Uchikata:2019frs}
\bibinfo{author}{\bibfnamefont{N.}~\bibnamefont{Uchikata}},
  \bibinfo{author}{\bibfnamefont{H.}~\bibnamefont{Nakano}},
  \bibinfo{author}{\bibfnamefont{T.}~\bibnamefont{Narikawa}},
  \bibinfo{author}{\bibfnamefont{N.}~\bibnamefont{Sago}},
  \bibinfo{author}{\bibfnamefont{H.}~\bibnamefont{Tagoshi}}, \bibnamefont{and}
  \bibinfo{author}{\bibfnamefont{T.}~\bibnamefont{Tanaka}}
  (\bibinfo{year}{2019}), \eprint{1906.00838}.

\bibitem[{\citenamefont{Mark et~al.}(2017)\citenamefont{Mark, Zimmerman, Du,
  and Chen}}]{Mark:2017dnq}
\bibinfo{author}{\bibfnamefont{Z.}~\bibnamefont{Mark}},
  \bibinfo{author}{\bibfnamefont{A.}~\bibnamefont{Zimmerman}},
  \bibinfo{author}{\bibfnamefont{S.~M.} \bibnamefont{Du}}, \bibnamefont{and}
  \bibinfo{author}{\bibfnamefont{Y.}~\bibnamefont{Chen}},
  \bibinfo{journal}{Phys. Rev.} \textbf{\bibinfo{volume}{D96}},
  \bibinfo{pages}{084002} (\bibinfo{year}{2017}), \eprint{1706.06155}.

\bibitem[{\citenamefont{Abedi and Afshordi}(2018)}]{Abedi:2018npz}
\bibinfo{author}{\bibfnamefont{J.}~\bibnamefont{Abedi}} \bibnamefont{and}
  \bibinfo{author}{\bibfnamefont{N.}~\bibnamefont{Afshordi}}
  (\bibinfo{year}{2018}), \eprint{1803.10454}.

\bibitem[{\citenamefont{Conklin et~al.}(2017)\citenamefont{Conklin, Holdom, and
  Ren}}]{Conklin:2017lwb}
\bibinfo{author}{\bibfnamefont{R.~S.} \bibnamefont{Conklin}},
  \bibinfo{author}{\bibfnamefont{B.}~\bibnamefont{Holdom}}, \bibnamefont{and}
  \bibinfo{author}{\bibfnamefont{J.}~\bibnamefont{Ren}} (\bibinfo{year}{2017}),
  \eprint{1712.06517}.

\bibitem[{\citenamefont{Tsang et~al.}(2018)\citenamefont{Tsang, Rollier, Ghosh,
  Samajdar, Agathos, Chatziioannou, Cardoso, Khanna, and Van
  Den~Broeck}}]{Tsang:2018uie}
\bibinfo{author}{\bibfnamefont{K.~W.} \bibnamefont{Tsang}},
  \bibinfo{author}{\bibfnamefont{M.}~\bibnamefont{Rollier}},
  \bibinfo{author}{\bibfnamefont{A.}~\bibnamefont{Ghosh}},
  \bibinfo{author}{\bibfnamefont{A.}~\bibnamefont{Samajdar}},
  \bibinfo{author}{\bibfnamefont{M.}~\bibnamefont{Agathos}},
  \bibinfo{author}{\bibfnamefont{K.}~\bibnamefont{Chatziioannou}},
  \bibinfo{author}{\bibfnamefont{V.}~\bibnamefont{Cardoso}},
  \bibinfo{author}{\bibfnamefont{G.}~\bibnamefont{Khanna}}, \bibnamefont{and}
  \bibinfo{author}{\bibfnamefont{C.}~\bibnamefont{Van Den~Broeck}},
  \bibinfo{journal}{Phys. Rev.} \textbf{\bibinfo{volume}{D98}},
  \bibinfo{pages}{024023} (\bibinfo{year}{2018}), \eprint{1804.04877}.

\bibitem[{\citenamefont{Cornish and Littenberg}(2015)}]{Cornish:2014kda}
\bibinfo{author}{\bibfnamefont{N.~J.} \bibnamefont{Cornish}} \bibnamefont{and}
  \bibinfo{author}{\bibfnamefont{T.~B.} \bibnamefont{Littenberg}},
  \bibinfo{journal}{Class. Quant. Grav.} \textbf{\bibinfo{volume}{32}},
  \bibinfo{pages}{135012} (\bibinfo{year}{2015}), \eprint{1410.3835}.

\bibitem[{\citenamefont{Littenberg and Cornish}(2015)}]{Littenberg:2014oda}
\bibinfo{author}{\bibfnamefont{T.~B.} \bibnamefont{Littenberg}}
  \bibnamefont{and} \bibinfo{author}{\bibfnamefont{N.~J.}
  \bibnamefont{Cornish}}, \bibinfo{journal}{Phys.Rev.}
  \textbf{\bibinfo{volume}{D91}}, \bibinfo{pages}{084034}
  (\bibinfo{year}{2015}), \eprint{1410.3852}.

\bibitem[{\citenamefont{Berti et~al.}(2006)\citenamefont{Berti, Cardoso, and
  Will}}]{Berti:2005ys}
\bibinfo{author}{\bibfnamefont{E.}~\bibnamefont{Berti}},
  \bibinfo{author}{\bibfnamefont{V.}~\bibnamefont{Cardoso}}, \bibnamefont{and}
  \bibinfo{author}{\bibfnamefont{C.~M.} \bibnamefont{Will}},
  \bibinfo{journal}{Phys. Rev.} \textbf{\bibinfo{volume}{D73}},
  \bibinfo{pages}{064030} (\bibinfo{year}{2006}), \eprint{gr-qc/0512160}.

\bibitem[{\citenamefont{Fisher}(1970)}]{fisher_method}
\bibinfo{author}{\bibfnamefont{R.~A.} \bibnamefont{Fisher}},
  \emph{\bibinfo{title}{Statistical methods for research workers. Fourteenth
  Edition Revised}} (\bibinfo{publisher}{Oliver and Boyd},
  \bibinfo{year}{1970}), ISBN \bibinfo{isbn}{0050021702}.

\bibitem[{\citenamefont{Abbott et~al.}(2019)}]{Abbott:2018wiz}
\bibinfo{author}{\bibfnamefont{B.~P.} \bibnamefont{Abbott}}
  \bibnamefont{et~al.} (\bibinfo{collaboration}{LIGO Scientific, Virgo}),
  \bibinfo{journal}{Phys. Rev.} \textbf{\bibinfo{volume}{X9}},
  \bibinfo{pages}{011001} (\bibinfo{year}{2019}), \eprint{1805.11579}.

\bibitem[{\citenamefont{Abbott et~al.}(2017{\natexlab{d}})}]{Abbott:2017dke}
\bibinfo{author}{\bibfnamefont{B.~P.} \bibnamefont{Abbott}}
  \bibnamefont{et~al.} (\bibinfo{collaboration}{LIGO Scientific, Virgo}),
  \bibinfo{journal}{Astrophys. J.} \textbf{\bibinfo{volume}{851}},
  \bibinfo{pages}{L16} (\bibinfo{year}{2017}{\natexlab{d}}),
  \eprint{1710.09320}.

\bibitem[{\citenamefont{Khanna and Price}(2017)}]{Khanna:2016yow}
\bibinfo{author}{\bibfnamefont{G.}~\bibnamefont{Khanna}} \bibnamefont{and}
  \bibinfo{author}{\bibfnamefont{R.~H.} \bibnamefont{Price}},
  \bibinfo{journal}{Phys. Rev.} \textbf{\bibinfo{volume}{D95}},
  \bibinfo{pages}{081501} (\bibinfo{year}{2017}), \eprint{1609.00083}.

\bibitem[{\citenamefont{Price and Khanna}(2017)}]{Price:2017cjr}
\bibinfo{author}{\bibfnamefont{R.~H.} \bibnamefont{Price}} \bibnamefont{and}
  \bibinfo{author}{\bibfnamefont{G.}~\bibnamefont{Khanna}},
  \bibinfo{journal}{Class. Quant. Grav.} \textbf{\bibinfo{volume}{34}},
  \bibinfo{pages}{225005} (\bibinfo{year}{2017}), \eprint{1702.04833}.

\end{thebibliography}

\end{document}